# Tuning the diffusion of magnon in $Y_3Fe_5O_{12}$ by light excitation

Shuanhu Wang[1], Gang Li[2], Er-jia Guo[3], Yang Zhao[1], Jianyuan Wang[1], Lvkuan Zou[4], Hong Yan[1], Jianwang Cai[2], Zhaoting Zhang[1], Min Wang[1], Yingyi Tian[1], Xiaoli Zheng[2], Jirong Sun[2*], Kexin Jin[1*]

1) Shanxi Key Laboratory of Condensed Matter Structures and Properties, School of Science, Northwestern Polytechnical University, Xi'an 710072, China
2) Beijing National Laboratory for Condensed Matter and Institute of Physics, Chinese Academy of Sciences, Beijing 100190, China
3) Neutron Science Directorate, Large Scale of Structure Group, Oak Ridge National Laboratory, Oak Ridge, TN 37831, USA.
4) High Magnetic Field Laboratory, Chinese Academy of Science, 230031 Hefei China
*e-mail: jrsun@iphy.ac.cn; jinkx@nwpu.edu.cn

## Abstract

Deliberate control of magnon transportation will lead to an energy-efficient technology for information transmission and processing. $Y_3Fe_5O_{12}$(YIG), exhibiting extremely large magnon diffusion length due to the low magnetic damping constant, has been intensively investigated for decades. While most of the previous works focused on the determination of magnon diffusion length by various techniques, herein we demonstrated how to tune magnon diffusion by light excitation. We found that the diffusion length of thermal magnons is strongly dependent on light wavelength when the magnon is generated by exposing YIG directly to laser beam. The diffusion length, determined by a nonlocal geometry at room temperature, is ~30 μm for the magnons produced by visible light (400-650 nm), and ~136-156 μm for the laser between 808 nm and 980 nm. The diffusion distance is much longer than the reported value. In addition to thermal gradient, we found that light illumination affected the electron configuration of the $Fe^{3+}$ ion in YIG. Long wavelength laser triggers a high spin to low spin state transition of the $Fe^{3+}$ ions in $FeO_6$ octahedron. This in turn causes a substantial softening of the magnon thus a dramatic increase in diffusion distance. The present work paves the way towards an efficient tuning of magnon transport behavior which is crucially important for magnon spintronics.

## Introduction

Magnons describe the deviation of a magnetic system from a fully magnetic order. Based on spin Seebeck effect (SSE)[1], non-equilibrium magnons can be generated by a thermal gradient across a magnet. The diffusion of these magnons forms magnon current or spin current. By injecting spin current into a heavy metal[1-3] or a topological insulator[4], which has strong spin-orbit coupling and thus a strong inverse spin Hall effect (ISHE), it can be converted into charge current that can be easily detected. The SSE has been observed in a wide range of materials, including ferrimagnetic[1,5,6], anti-ferromagnetic[7] and even some paramagnetic materials[8]. However, the diffusion length of the non-equilibrium magnon is still in debate. A recent report showed that the SSE first increased and then was saturated as the YIG film thickness increased[9]. At room temperature, the characteristic thickness for the SSE-saturation was ~0.1 μm for pulsed laser deposited YIG. Using laser illumination to locally break the thermal equilibrium between magnons and phonons, which can be probed directly by micro-Brillouin light scattering, An *et al.*[10] found that the magnon diffusion length is 3.1 μm at about 372 K.

Recently, nonlocal spin Seebeck geometry[11-16] was widely adopted to investigate spin transport behavior; the spatially separated structure makes the measurements immune to parasitic thermoelectric effect[17,18]. Through a spin accumulation in Pt in a nonlocal structure, the magnons can be generated and detected in a separated structure. The first work in this aspect was done by Cornelissen *et al.* in 2015[11]. The authors found that the diffusion length of the non-equilibrium magnon is ~9.4 μm at room temperature for the YIG film with thickness of 200 nm. Focusing laser spot on a Pt absorption pad to generate thermal magnons and detecting the ISHE signals from a Pt bar separated from the laser spot[12,16], Giles *et al.* declared that the magnon diffusion length is less than 9 μm at room temperature and at least 47 μm at 23 K[12].

Compared to simply determining diffusion distance, tuning the transport behavior of the magnon is obviously much more important and challenging. Despite the tremendous advances on spincalorics, works in this regard are rare. It is well known that the transport behavior of the magnon strongly depends on the dispersion relation. Earlier investigations on optical spectra showed that the electron configuration of the $Fe^{3+}$ ions in YIG can be affected by light excitation[19, 20]. This in turn will affect the dispersion relation of the magnon, due to the change in spin state and superexchange interaction between $Fe^{3+}$ ions. These works provide us the suggestion of tuning the magnon diffusion by light excitation rather than the conventional opto-thermal and electro-thermal techniques. Based on a specifically designed nonlocal geometry, we performed a systematic investigation on transport behavior of the magnon in YIG, focusing on tuning magnon diffusion by laser illumination. Exposing the sample directly to a laser spot, we find diffusion length of the magnon is strongly dependent on the wavelength of excitation light. The diffusion distance is ~30 μm when induced by visible light (405-650 nm), and ~137-156 μm when light wavelength is between 808 nm and 980 nm. Therefore, the laser excitation causes not only thermal gradient but also induces the electron configuration transition of the $Fe^{3+}$ ions in YIG, and then modifies the dispersion relation and the magnon diffusion length.

## Experiment

The YIG films were grown on (111)-oriented $Gd_3Ga_5O_{12}$ (GGG) substrates (5×3×0.5 mm$^3$) by the techniques of pulsed laser deposition (PLD) and liquid phase epitaxy (LPE), respectively, with the corresponding film thicknesses of 40 nm and 20 μm. Then a Pt layer with a thickness of 5 nm was deposited by magnetron sputtering on YIG through a bar-shaped mask. The size of Pt strip was 4.8×0.5 mm$^2$. The surface morphology of YIG was measured by atomic force microscopy (Supplementary Fig. S1), which shows a root mean square roughness of 1.2 Å (PLD sample) or 1.1 nm (LPE sample). Smooth surface is expected to favor a high spin mixing conductance at the Pt-YIG interface[21]. X-ray diffraction (XRD) analysis confirmed the epitaxial growth of YIG on GGG and the high film quality, as indicated by the sharp (444) reflection and the appearance of interference peaks (Supplementary Fig. S2). The film thickness of YIG (PLD sample) and Pt was determined by low-angle X-ray reflectivity. Further details on sample preparations and characterizations can be found in Supplementary materials.

Magnetic field ($H$) was provided by two Helmholtz coils, applied along the $x$-axis of the sample. Two electrodes aligning along the $y$-axis were used to detect the ISHE voltage ($V_{ISHE}$). The sample was attached to cryostat by silver paste to get good thermal contact and to absorb the transmitted thermal energy from the laser beam. The cryostat is sealed in an electromagnetically shielding box with an optical window. The laser beam with a preset wavelength was focused, though a convex mirror, on sample surface to generate an up-to-bottom thermal gradient. The diameter of the light spot was less than 20 μm, measured through an infrared macro lens. As shown in Fig. 1a, the laser and the convex mirror were mounted on a lead rail along the $x$-axis, which allows a position tuning in micrometers. The output voltage across the Pt bar was recorded as the laser spot sweeping through the middle of the Pt bar and the surface of the YIG film, along the $x$-axis. The ISHE voltage is calculated by $V_{ISHE}=[V_{ISHE}(+H)-V_{ISHE}(-H)]/2$, where $V_{ISHE}(+H)$ and $V_{ISHE}(-H)$ are the saturation voltages in two oppositely directed magnetic fields. The maximal magnetic field in this experiment is 120 Oe. It is so small that the influence of the opening Zeeman gap on the thermal magnon can be ignored[22]. For clarity, the $V_{ISHE}$ induced by the laser with the wavelength of λ nm was noted as $V_{ISHE}^{\lambda}$.

## Results and discussions

When the top surface of the sample is illuminated by a laser beam, the YIG film will absorb a part of the energy. For the film deposited by LPE, the transmitted energy is very low. So the absorbed power can be directly determined by simultaneously measuring the incident and reflected powers by optical power meters. In general, the absorbed energy will establish an out-of-plane thermal gradient, generating thermal magnons due to the SSE. The thermal magnons will then diffuse laterally towards the Pt bar, yielding an electrical voltage ($V_{ISHE}$) due to the ISHE. Although an in-plane thermal gradient could also be produced by the absorbed energy, it will mainly locate at the region of the laser spot and decays rapidly in the lateral direction. As proven by the results of finite-element model (FEM) simulation conducted by Giles *et al*[12] and An *et al.*[10], the temperature of the YIG surface will return to ambient temperature within 15 μm. Therefore, lateral heat flow should have no detectable effect on the measurement of diffusion length.

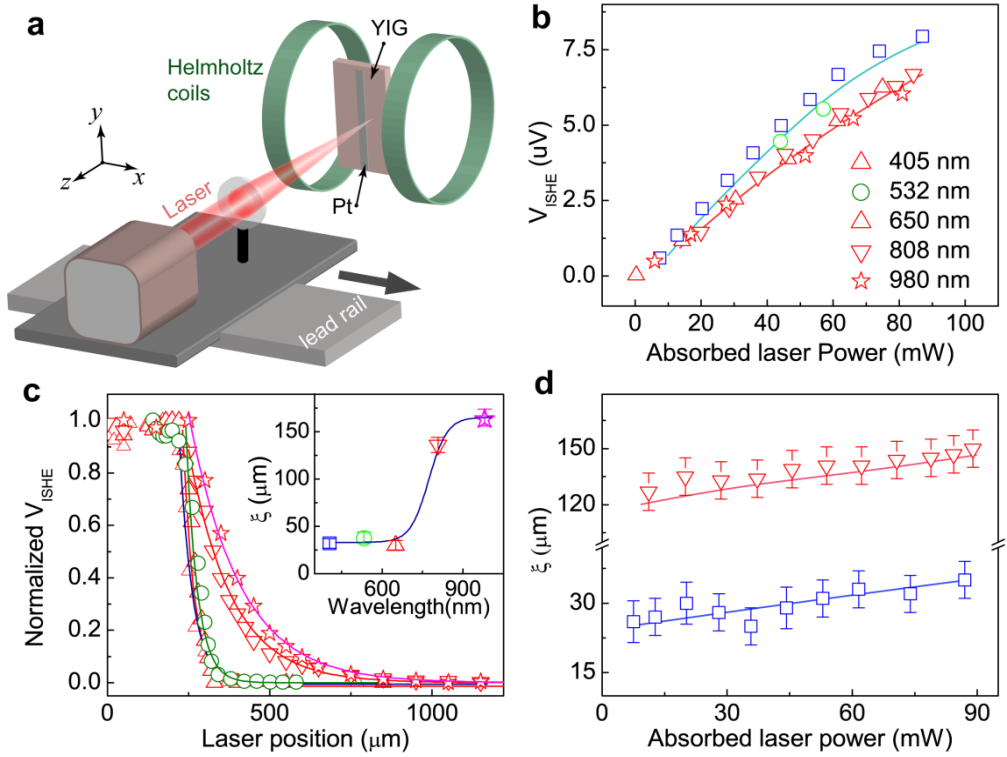

**Figure 1 | Dependence of magnon diffusion length on laser wavelength. a,** The schematic diagram of the experimental setup. **b**. $V_{ISHE}$ as a function of the absorbed laser power, measured with the lasers of different wavelength. Laser spot was positioned at the middle of the Pt bar (x=0). **c**. Normalized $V_{ISHE}$ as a function of spot position. Symbols represent the experimental data and solid lines are the results of curve fitting based on Eq. (1). Inset plot shows the diffusion length $\xi$ as a function of laser wavelength, deduced from data fitting. **d**. Dependence of the diffusion length on absorbed laser power. The error bars represent the standard error in the fits. All measurements were conducted at room temperature. Only the data for the YIG prepared by LPE are shown here. The solid lines in **b, d** and inset of **c** are guides to the eyes.

Fig. 1a is a sketch of the experiment setup. Fig. 1b shows the $V_{ISHE}$ as a function of the absorbed laser power. When the laser beam is focused on the middle of the Pt bar, $V_{ISHE}$ linearly increases with laser power as expected. When covering the whole YIG with a Pt layer (5 nm in thickness), we found that $V_{ISHE}$ remained constant, regardless of the location and size of the laser spot (Supplementary Fig. S2). $V_{ISHE}^{405}$ and $V_{ISHE}^{532}$ are a slightly larger than $V_{ISHE}^{808}$ and $V_{ISHE}^{980}$. This can be ascribed to the larger absorption coefficient of lights of 405 nm and 532 nm[19].

Fig. 1c presents the dependence of the normalized $V_{ISHE}$ on the position of laser spot. Setting the middle of the Pt electrode to *x*=0, and collecting $V_{ISHE}$ as laser spot sweeps along *x*-axis, we found that the $V_{ISHE}$ kept nearly constant when laser spot scans across the Pt strip (-0.25 mm<x<0.25 mm), and exponentially decreased with the distance away from the Pt bar. Fascinatingly, the decay rate strongly depends on wavelength. Illuminated by visible light (405 nm, 532 nm and 605 nm), $V_{ISHE}$ drops to zero immediately when the light spot moves out of the region of the Pt bar. In contrast, it decays slowly with the distance from Pt for the light of 808 nm and 980 nm, remaining sizable when laser spot is 0.5 mm away from the Pt edge. A further analysis indicates that the $V_{ISHE}^{\lambda}$-*x* relation can be well described by

$$V_{ISHE}^{\lambda} = V_0^{\lambda} \exp\left(-\frac{x}{\xi^{\lambda}}\right) \quad (1)$$

where $V_0^{\lambda}$ is normalized coefficient and $\xi^{\lambda}$ is the diffusion length of the magnons induced by the laser of *λ* nm. The inset plot in Fig. 1c depicts the magnon diffusion distance as a function of laser wavelength. The thermal magnons induced by long wavelength lasers transport much longer distance than the ones induced by short wavelength lights. The diffusion lengths $\xi^{405}$, $\xi^{532}$ and $\xi^{650}$ are all around 30 μm while $\xi^{808}$ is ~137 μm and $\xi^{808}$ is 156 μm. The diffusion length is not sensitive to laser power as shown in Fig. 1d. There are reports on the diffusion length of thermal magnon determined by different techniques, showing that $\xi$ is generally shorter than several micrometers[10-13]. The long diffusion length and its strong wavelength dependence observed here indicate that the light not only acts as a heating source to generate thermal gradient but also affects, in some way, the characteristics of the thermal magnon, assigning the latter unconventional diffusion behaviors.

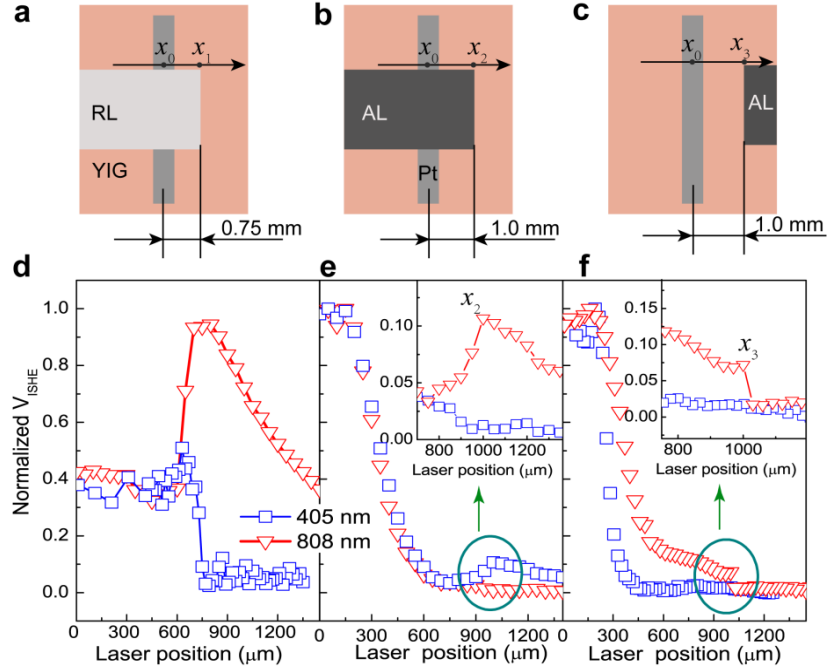

**Figure 2 | Experiment for deep study of the wavelength dependent magnon diffusion length. a,** sample is covered by polished aluminum foil (note as RL) to reflect the injected laser. **b** and **c,** an absorbed layer (AL) is attached on the sample. $x_0$ is the middle of the sample, while $x_1$ and $x_2$ are the right side of RL and AL as noted in **a** and **b** respectively. $x_3$ is the left side of AL in **c**. **d, e** and **f,** are the $V_{ISHE}$ detected in the Pt electrode when laser spot is scanned from x=0 on the surface of sample designed as **a. b.** and **c.** respectively. Insets of **e** and **f** are the enlarge view of green circle region to show the details of the curve change. The absorbed laser power is set to 30 mW for both lasers.

To distinguish the effects of thermal gradient and photo excitation on magnon diffusion, we selected a short wavelength (405 nm) and a long wavelength (808 nm) laser for further investigations. As show in Fig. 2a, we first covered a part of the Pt bar and the YIG film by an aluminum foil, which acts as reflected layer (RL) for incident light, and then repeated the experiments (to avoid short circuiting, an insulating sheet was inserted between RL and Pt). When the laser is focused on the region covered by the RL, a weak $V_{ISHE}$ is observed. It can be ascribed to the simple thermal effect; absorption of the incident light by the RL caused a temperature growth of the RL and then a thermal gradient underneath YIG, driving a magnon current. Due to the high thermal conductivity of the RL, the thermal gradient could be uniform in YIG. Therefore, $V_{\text{ISHE}}$ is independent of the position of the laser spot. When the light spot moves out of the right edge of the RL (x=0.5 mm), however, $V_{ISHE}^{405}$ drops immediately to zero but $V_{ISHE}^{808}$ jumps to a higher value before a slow exponential decay with the distance from the Pt bar. The undetectable $V_{ISHE}^{405}$ means that the corresponding magnons are unable to reach the Pt bar by diffusion when they are 0.5 mm away from the right side of Pt bar. However, the magnons excited by the light of 808 nm can transport an obviously longer distance than 0.5 mm, and then they are collected by Pt. This result indicates that the thermal effects of the two lights are similar but the excitation effects are different.

In the above paragraph, the irradiated RL generated a uniform thermal gradient. We also studied the effects of local thermal gradient on the magnon diffusion length. As shown in Fig. 2b, a black insulating absorb layer (AL) was used to cover the Pt bar and the YIG film. The absorbed energy by AL can be directly transferred to the YIG film when AL is illuminated, generating a $V_{\text{ISHE}}$ as shown in Fig. 2e. Since the thermal conductivity of the AL is low, the maximal thermal gradient will appear exactly underneath the laser spot. As expected, $V_{\text{ISHE}}$ is the highest at $x$=0, and rapidly decreases as laser spot moves along $x$-axis. Notably, the short and long wavelength lasers produce similar $V_{\text{ISHE}}$-$x$ dependences as long as the laser spot locates on AL. This result implies that the thermal magnons produced by thermal gradient are the same in nature, regardless of the heating source. Here $V_{ISHE}^{405}$ does not sharply drop to zero like in Fig. 1c when the laser spot leaves the Pt bar, because the thermal energy will laterally expand inside AL, broadening the region with thermal gradient. Fascinatingly, as soon as the light spot completely leaves the AL, a sudden jump happens to $V_{ISHE}^{808}$, whereas $V_{ISHE}^{405}$ keeps zero without any anomalous variation. Placing the AL 1 mm away from the Pt (Fig. 2c), we observed null $V_{\text{ISHE}}$ as laser spot sweeps through the AL, meaning that the thermally excited magnon cannot reach the position of the Pt bar. However, when the laser spot gets out of the left edge of the AL, $V_{ISHE}^{808}$ becomes nonzero immediately while $V_{ISHE}^{405}$ remains zero until the light spot is close enough to Pt. Again, long wavelength light shows its advantage over the short one in generating long-diffusion-distance magnons. Based on the above results, we come to the conclusion that the diffusion length of thermal magnon generated by infrared light is much longer than the one induced by visible light and also the one by a simple thermal gradient.

As a supplement, we would like to point out that ultra-long diffusion length $\xi^{808}$ has nothing to do with in-plane thermal gradient. Because the absorbed energy is identical for both lasers (30 mW), it will generate the same lateral thermal gradient. And if the $V_{ISHE}$-$x$ relation is determined by the lateral thermal gradient, it will result the same spatial evolution of $V_{ISHE}^{808}$ and $V_{ISHE}^{405}$ which is obviously not the case in Fig. 2.

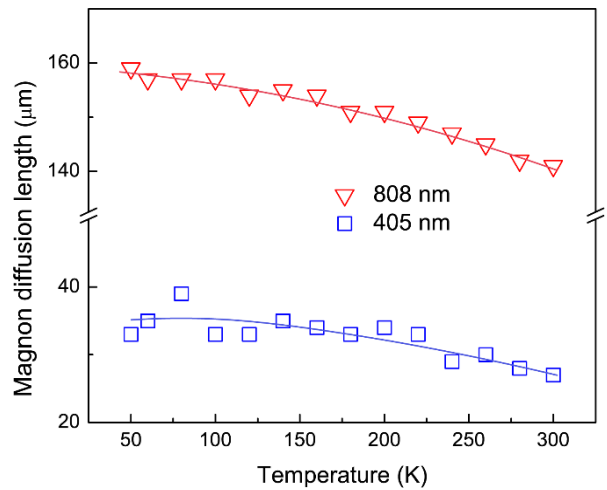

**Figure 3 | Temperature dependence of thermal magnon diffusion length.** The laser power is fixed at 30 mW. The solid lines are guides to the eyes.

The temperature dependence of the diffusion length of magnons is shown in Fig.3. According to the Bose-Einstein distribution, the amount of thermal magnons and phonons will decrease with decreasing temperature. This means that the scattering between magnon and phonon will be weakened, and the diffusion distance will be increased. Guo et al.[23] found that $\xi \propto T^{-1}$, i.e., diffusion length is closely related to the number of phonons and magnons. On the contrary, a pervious study with nonlocal geometry found a slight reduction of $\xi$ with decreasing temperature, and ascribed this phenomenon to a compensation effect of increased relaxation time by reduced thermal velocity of the magnon. However, Fig. 3 shows that both $\xi^{405}$ and $\xi^{808}$ are only slightly increased upon cooling, rather than proportional to $T^{-1}$ or slightly decreasing. The different $\xi$-$T$ dependence in Fig. 3 implies that the diffusion behavior of the non-equilibrium magnon induced by light irradiation is unique.

We also performed the same investigations for the thinner YIG film prepared by the PLD technique (40 nm). We found the diffusion length in PLD-prepared YIG is a bit larger than that in LPE-prepared YIG (Supplementary Fig. S4). For example, $\xi^{808}$ will reach up to 162 μm which is due to the much smoother surface which decrease the scatter during the lateral transport of magnon. But the diffusion length in the two sample is comparable to each other which is consistent with the report that the magnon transport along film plane is independent of film thickness[13]. The diffusion length in the PLD-prepared YIG is also strongly dependent on the excitation laser which indicates that the difference in $\xi$ induced by different lights is a general feature for YIG.

Our observation cannot be ascribed to the so-called photo-spin-voltaic effect[24]. A recent study showed that when the Pt/YIG hybrid structure is exposed to light, especially infrared light, a photon-driven spin-dependent electron excitation will occur near the Pt-YIG interface, producing a photo-spin-voltaic effect. Since this effect is independent of the direction of temperature gradient, reversing the direction of thermal gradient will not change its sign. We reversed the incident direction of the light (λ=808 nm) and found a sign change of $V_{ISHE}$ (Supplementary Fig. S5). Illuminating the back side of the sample, moreover, we observed the same magnon diffusion length as illuminating the front side (Supplementary Fig. S5). All these show that the $V_{ISHE}$ detected here is unambiguously generated by non-equilibrium magnons instead of the photo-spin-voltaic effect.

To explore the mechanism of the wavelength dependent magnon diffusion length, we measured the absorption spectrum of YIG as shown in Fig. 4. According to earlier researches[21, 28], the peak around 800-1000 nm in the absorption spectrum arising from the $^6A_{1g}(^6S) \rightarrow {}^4T_{1g}(^4G)$ transition of the electron configuration in octahedral crystal field[19, 20], while the 650, 532, and 405 nm are corresponding to the $^6A_1(^6S) \rightarrow {}^4T_1(^4G)$, $^6A_1(^6S) \rightarrow {}^4T_2(^4G)$, and $^6A_1(^6S) \rightarrow {}^4T_2(^4D)$ transitions in the tetrahedral crystal field, respectively. Visible light may also be absorbed by the electron in octahedral crystal field. However, due to the octahedral symmetry, transitions between the ground state and any excited states are parity forbidden. As a result, the visible light is nearly totally absorbed by the electron in tetrahedral crystal field. It is well known that two of the five $Fe^{3+}$ ions in YIG are octahedrally coordinated (*a* sites) while other three are tetrahedrally coordinated (*d* sites), forming two sublattices. The magnetic moments of the two sublattices arrange in antiparallel, i.e., the YIG is ferrimagnetic. The magnetic moment of the *a*-site (*d*-site) $Fe^{3+}$ is antiparallel (parallel) to external magnetic field as well as the net magnetization.

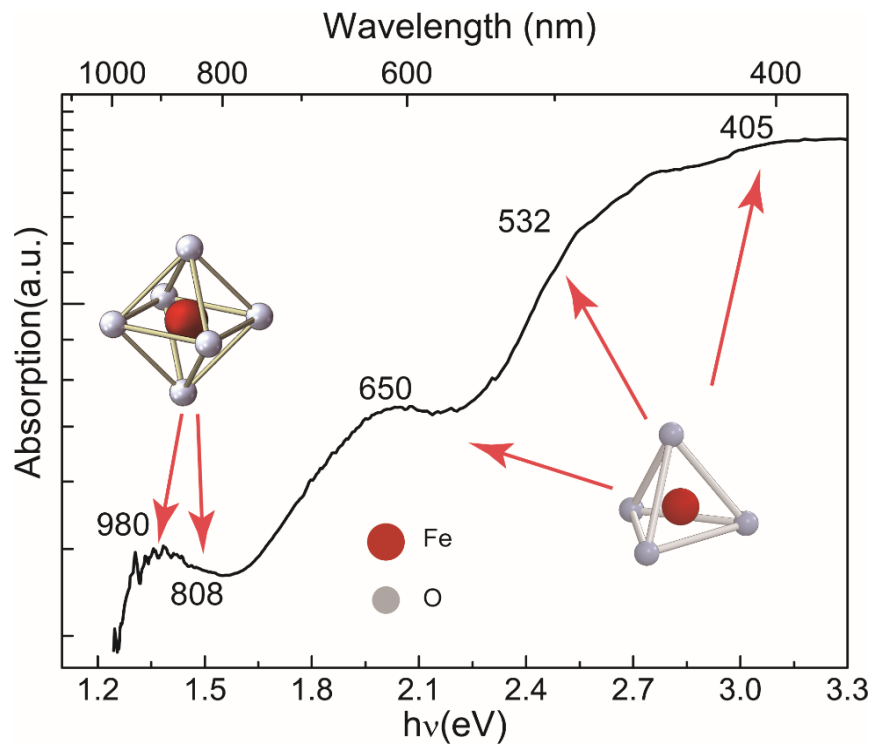

**Figure 4 | Absorption spectrum of YIG prepared by PLD.**

To reveal the underlying physics of the unusual diffusion behavior of the photo excited magnons, we performed a simple analysis of the factors that affect the kinetic behavior of the magnons. Ignoring the influence of $J_{aa}$ and $J_{dd}$ (superexchange constant $J_{ad}$ is much larger than $J_{aa}$ and $J_{dd}$[25]), we have the dispersion relation of the acoustic branch spin wave[26]

$$\hbar\omega_k = 4|J_{ad}|a^2 \frac{S_d S_a}{S_d - S_a} k^2, \tag{2}$$

for ferrimagnetic material with cubic crystal system under long wavelength approximation, where $a$, $\omega_k$ and $k$ are respectively lattice constant, the frequency and wave vector of the magnon. $S_a$ and $S_d$ are the spin quantum number of the $Fe^{3+}$ ions on $a$ and $d$ sites. Obviously, the relaxation process will cause a dissipation of the magnons. The relaxation time $\tau$ has been calculated in many works[27, 28]. Following the procedure in Ref[27], it can be proven $\tau \sim D^{-1}$, where $D=4|J_{ad}|a^2 \frac{S_d S_a}{S_d - S_a}$, in this case, is the stiffness constant of the magnon. There is a simple relation between diffusion length and the stiffness constant $\xi \propto D^{-0.5}$, adopting the relation $\xi \propto \tau^{0.5}$. Therefore, if the variation of $D$ with the electron configuration transition is known, the diffusion behavior of the magnons would be understood.

A simple analysis shows that a direct consequence of the $^6A_{1g} \to {}^4T_{1g}$, $^6A_1 \to {}^4T_1$, , and $^6A_1 \to {}^4T_2$ switching is spin state transition; the number on the upper left corner of each symbol is 2S+1, with S being the spin angular quantum number of $Fe^{3+}$. The ground state and the excited state are a high-spin state and a low-spin state, respectively. When the spin state of the $Fe^{3+}$ ion at $a$ site ($d$ site) changes from high spin state to low spin state under illumination, it will cause a reduction of $S_a$ ($S_d$) from 5/2 to 3/2, thus a decrease (increase) in $\frac{S_d S_a}{S_d - S_a}$ or, equivalently, a softening (hardening) of the magnons. In addition to $\frac{S_d S_a}{S_d - S_a}$, $J_{ad}$ is also a factor affecting the dispersion relation. According to theory of superexchange[29], $J_{ad}$ is proportional to $\Delta^{-2}$, where $\Delta = E_d - E_p$, is the energy difference between the 3d and 2p orbital states of the magnetic ion and the oxygen ion, respectively. The energy splitting of the $3d^5$ state is strongly dependent on the configuration of the surrounding oxygen ions. $E_d$ will grow when the electron transfer from the $^6A_{1g}(^6S)$ to the $^4T_{1g}(^4G)$ configuration, causing a decrease in $J_{ad}$. For the tetrahedral crystal field, according to Eq. (2) the effects of the spin state transition and the variation of $J_{ad}$ counteract each other. For the octahedral crystal field, on the contrary, the two effects enhance each other, making the magnons softened.

From the absorption spectrum in Fig. 4, the long-wavelength lights only affect octahedral site $Fe^{3+}$ ($a$ site) while the visible light mainly influence tetrahedral site $Fe^{3+}$ ($d$ site). Based on the above analysis, the former process produces softer magnons than the latter one. According to the relation $\xi \propto D^{-0.5}$, soft magnons will have a long diffusion distance. Therefore, the magnon diffusion can be selectively tuned by changing the electron configuration of $Fe^{3+}$ ion.

It is worth mentioning that the lateral magnon diffusion length deduced from the nonlocal geometry could be substantially different from the diffusion length derived from local configuration. In the latter case the spin current is parallel to temperature gradient, thus suffers from the scattering of surface and interface[9, 23]. As proven by recent investigations, the SSE mainly originated from long-wavelength magnons[30, 22]. Therefore, the magnons detected here are mainly long-wavelength ones which present lower relaxation rate[31]. This is one of the reasons why the diffusion distance here is longer than that determined by other techniques such as Brillouin scattering which takes thermal magnons of the whole spectrum into account[9]. We also noticed the experiment conducted by Gile *et al*.[12, 16] The authors use a Pt pad to generate thermal gradient and reported a diffusion length of 10 μm at 250 K. As demonstrated by our experiments (Fig. 2), the diffusion length is very short for magnons induced by thermal gradient alone, which is a conclusion consistent with that of Gile *et al*. Of cause, the incident light in the work of Gile *et al*. maybe also transmitted to YIG through the Pt pad. However, the Pt pad is 10 nm in thickness, and the transmitted intensity could be too weak to produce any significant effects on the electron configuration.

## Conclusion

In summary, we experimentally show that the diffusion length of the non-equilibrium magnon in YIG is strongly dependent on laser wavelength when they are induced by laser irradiation. Magnons generated by the laser beams of 808 nm and 980 nm can be detected from a distance as far as 1 mm. The corresponding diffusion lengths are 137 μm and 156 μm, respectively, while the diffusion lengths of the magnons excited by visible light (400-650 nm) is only ~30 μm. We found unambiguous correspondences between the electron configuration and magnon diffusion. Long-wavelength laser excites a transition of the electron configuration for the $FeO_6$ octahedron in YIG and induces softened magnons which have a longer diffusion distance. In contrast, visible light only affects electron configuration in the $FeO_4$ tetrahedron, and the corresponding magnons are stiff. The present work demonstrates for the first time the tuning of magnon diffusion by selective modification of electron configuration. The principle proven here can be extended to other materials.

## Acknowledgements

This work is supported by the National Natural Science Foundation of China (Nos. 11604265, 51402240, 11520101002 and 51572222) and the Fundamental Research Funds for the Central Universities (No. 3102017jc01001). J.R. Sun thanks the support of the National Basic Research of China (2016YFA0300701) and Key Program of the Chinese Academy of Sciences.

## Author contributions

S.H. Wang conceived the idea, established the experimental setup and analyzed the data. G. Li, Y.Y. Tian, X.L. Zheng and L.K. Zou fabricated and characterized the sample together. S.H. Wang, J.Y. Wang, H. Yan, Z.T. Zhang and M. Wang performed the measurement. J.R. Sun, S.H. Wang, K.X. Jin, E.J. Guo, Y. Zhao and J.W. Cai provide the theoretical analysis. S.H. Wang and J.R. Sun wrote the paper and the supplementary information with help from all the other co-authors.